\documentclass[twocolumn]{jpsj3} %% two-column layout
\title{Effect of K Doping on Phonons in Ba$_{1-x}$K$_x$Fe$_2$As$_2$}

\author{Chul-Ho LEE$^{1,2}$, Kunihiro KIHOU$^{1,2}$, Kazumasa HORIGANE$^3$, Satoshi TSUTSUI$^4$, Tatsuo FUKUDA$^{2,5,6}$, Hiroshi EISAKI$^{1,2}$, Akira IYO$^{1,2}$, Hirotaka YAMAGUCHI$^1$, Alfred Q. R. BARON$^{2,4,5}$, Markus BRADEN$^7$, and Kazuyoshi YAMADA$^3$}

\inst{$^{1}$National Institute of Advanced Industrial Science and Technology, Tsukuba, Ibaraki 305-8568, Japan\\
$^{2}$JST, Transformative Research-Project on Iron Pnictides (TRIP), Chiyoda, Tokyo 102-0075, Japan\\
$^{3}$WPI, Tohoku University, Sendai 980-8577, Japan\\
$^{4}$Japan Synchrotron Radiation Research Institute, SPring-8, Sayo, Hyogo, 679-5198, Japan\\
$^{5}$Materials Dynamics Laboratory, RIKEN SPring-8 Center, Sayo, Hyogo 679-5148, Japan\\
$^{6}$JAEA, SPring-8, Sayo, Hyogo 679-5148, Japan\\
$^{7}$$I\hspace{-.1em}I$. Physikalisches Institut, Universit\"at zu K\"oln, Z\"ulpicher Str. 77, D-50937 K\"oln, Germany}

\abst{The lattice dynamics of Ba$_{1-x}$K$_x$Fe$_2$As$_2$ ($x$ =
0.00, 0.27) have been studied by inelastic X-ray scattering measurement at
room temperature.  K doping induces the softening and 
broadening of phonon modes in the energy range $E$ = 10-15 meV.  
Analysis with a Born-von K\'{a}rm\'{a}n force-constant
model indicates that the softening results from reduced
interatomic force constants around (Ba,K) sites following the
displacement of divalent Ba by monovalent K. The phonon broadening
may be explained by the local distortions induced by the K
substitution. Extra phonon modes are observed around the wave
vector $\bf q$ = (0.5,0,0) at $E$ = 16.5 meV for the $x$ = 0.27
sample. These modes may arise either from the local disorder
induced by K doping or from electron-phonon coupling.}

\kword{Fe-based superconductor, inelastic X-ray scattering, phonon dispersion}

\begin{document}
\maketitle

\section{Introduction}

Since the discovery of Fe-based superconductors with superconducting transition temperatures ($T_c$) of 
up to 55 K \cite{Kamihara2008,Ren-F-2008,Kito2008,Ren-O-2008, Miyazawa2009}, intensive studies 
have been conducted to clarify the mechanism of Cooper pair formation and to improve their $T_c$ values.  
For example, the possibility of phonon-mediated superconductivity has been studied intensively \cite{Litvinchuk2008,Fukuda2008,Christianson2008, Mittal2008, Zbiri2009, Mittal2009, Tacon2008, Reznik2009, Boeri2008,Noffsinger2009}.  
Calculations using the density functional perturbation theory, however, 
revealed very weak electron-phonon coupling constants, 
suggesting that, within those simplified models, 
conventional phonon-mediated superconductivity is unlikely \cite{Boeri2008,Noffsinger2009}.  
Nevertheless, a  mechanism involving phonons remains possible. 
In fact, both $T_c$ and the band structure of Fe-based superconductors are sensitive to 
the crystal structure \cite{Lee2008,Singh2008,Kuroki2009}, 
suggesting the importance of electron-lattice coupling.  
For example, the coupling between lattice and magnetism \cite{Yildirim2009} or Fermi surface nesting \cite{Eschrig2009} may induce high-$T_c$ superconductivity.  
Recently, the results of neutron scattering measurements on CaFe$_2$As$_2$ have been 
interpreted as evidence for strong electron-phonon coupling \cite{Mittal2009}.  
Inelastic X-ray scattering measurements on NdFeAsO$_{1-y}$F$_y$ powder samples revealed softening in the phonon density of states under F doping \cite{Tacon2008}.  
In contrast, recent studies on doped and undoped BaFe$_2$As$_2$ single crystals 
by inelastic X-ray scattering measurement revealed no
significant doping-induced phonon softening  \cite{Reznik2009}.  
Further studies on phonon dynamics using single crystals are essential for 
identifying relevant phonon modes and elucidating the role of phonons 
in the appearance of superconductivity in Fe-based superconductors.  

%=========================================================
\begin{figure}
\includegraphics[width=\columnwidth]{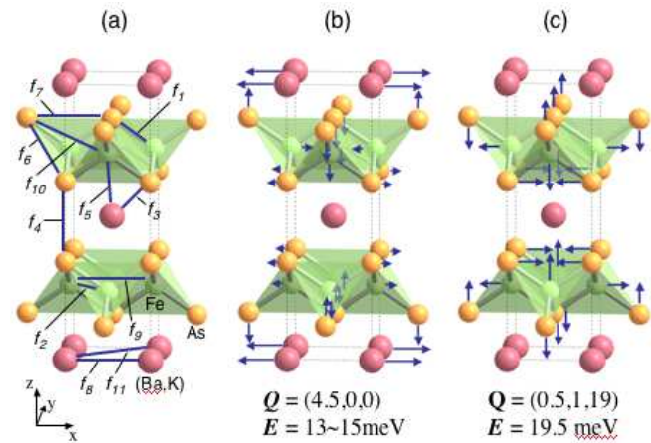}
\caption{\label{structure} Crystal structures of Ba$_{1-x}$K$_x$Fe$_2$As$_2$ with (a) labeled interatomic force constants, and instantaneous displacement patterns of phonon modes at (b) $\bf Q$ = (4.5,0,0) and (c) $\bf Q$ = (0.5,1,19).  (Figs. \ref{spectrum-long} and \ref{spectrum-trans})}
\end{figure}
%=========================================================
%=========================================================
\begin{figure}
\includegraphics[width=\columnwidth]{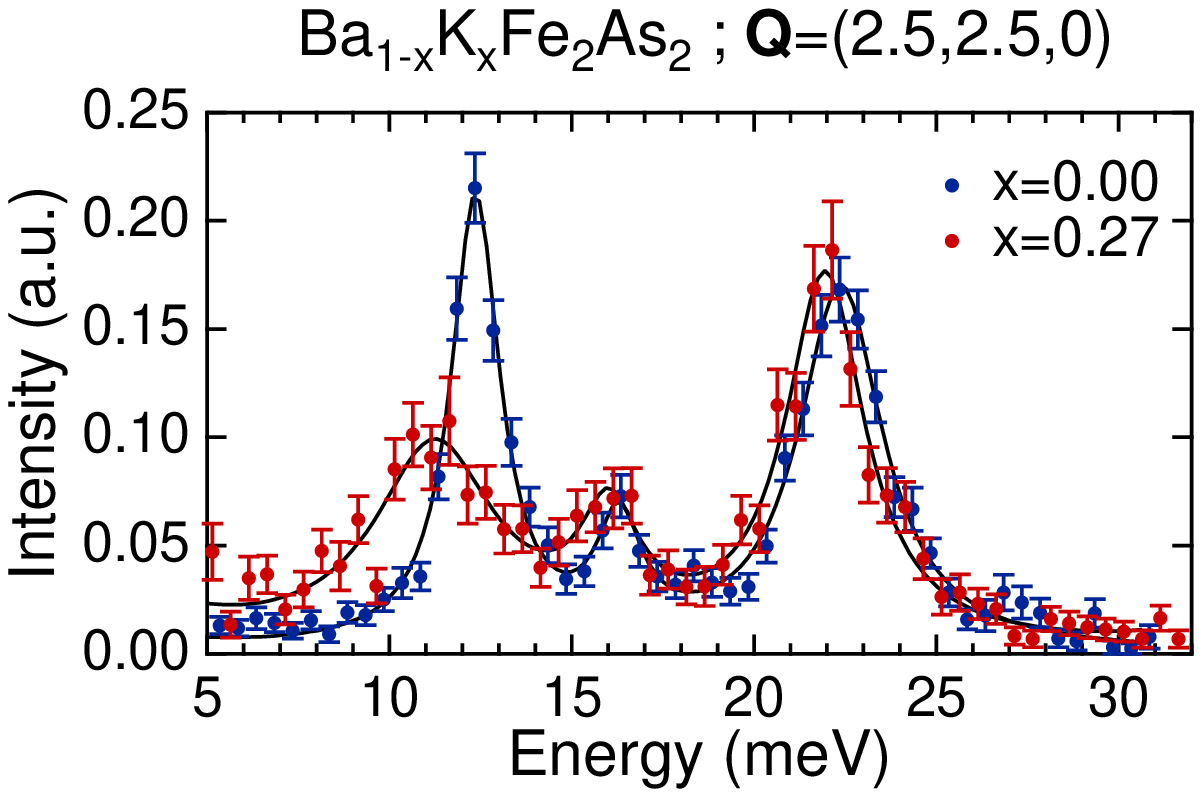}
\caption{\label{spectrum-nest} Energy spectra of phonons observed at $\bf Q$ = (2.5,2.5,0) for Ba$_{1-x}$K$_x$Fe$_2$As$_2$ ($x$ = 0.00, 0.27).  Solid lines show the results of fits with three Lorentzian components.}
\end{figure}
%=========================================================

%=========================================================
\begin{figure}
\includegraphics[width=\columnwidth]{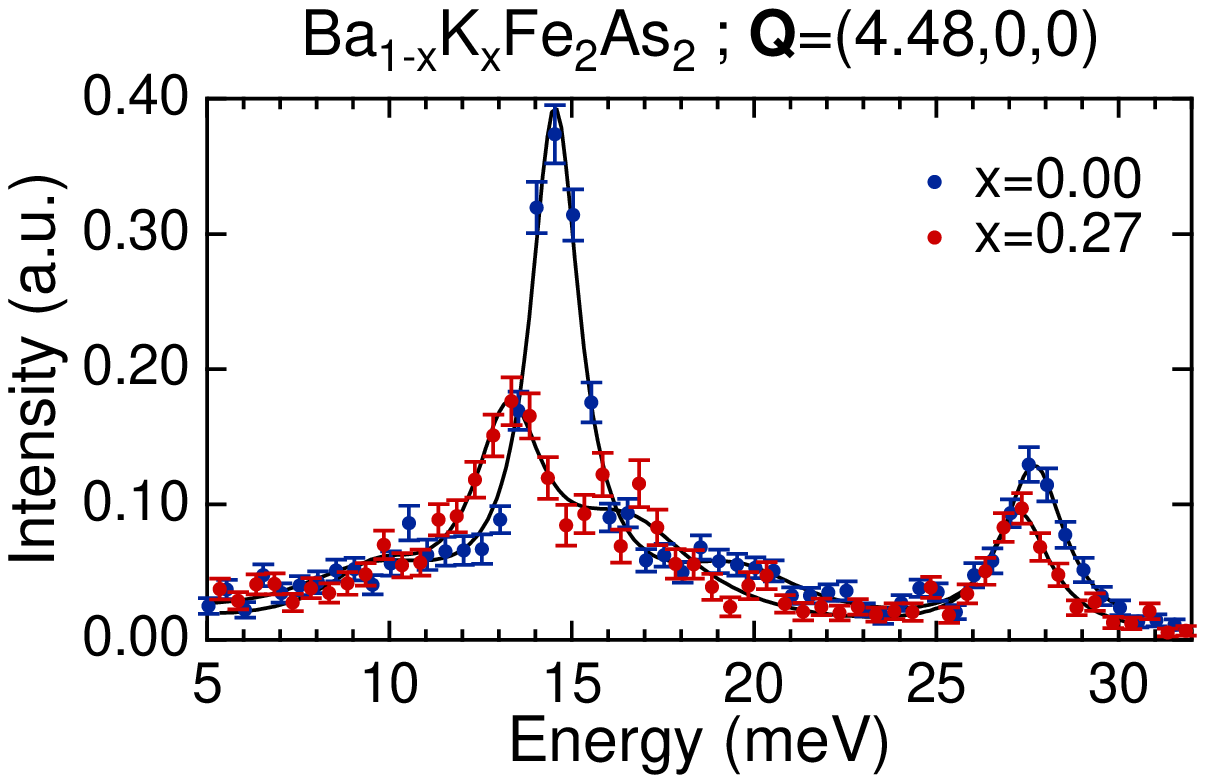}
\caption{\label{spectrum-long} Energy spectra of phonons observed at $\bf Q$ = (4.48,0,0) for Ba$_{1-x}$K$_x$Fe$_2$As$_2$ ($x$ = 0.00, 0.27).  Solid lines show the results of fits with four Lorentzian components.}
\end{figure}
%=========================================================
%=========================================================
\begin{figure}
\includegraphics[width=\columnwidth]{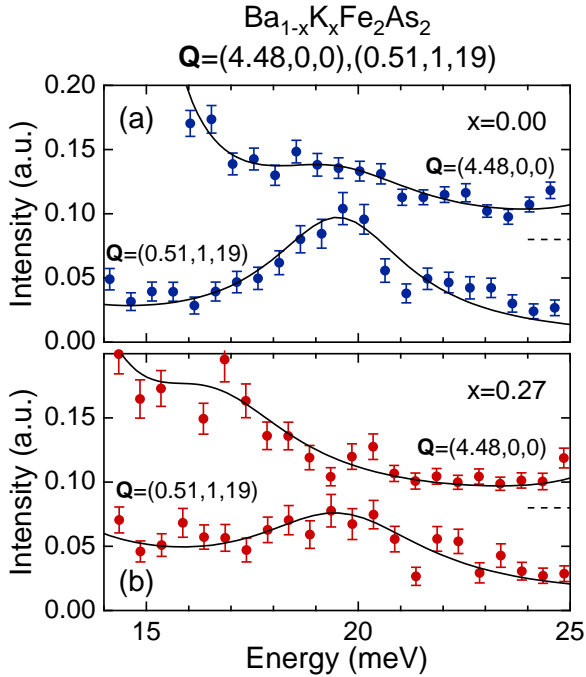}
\caption{\label{spectrum-trans} Energy spectra of phonons observed at $\bf Q$ = (4.48,0,0) and (0.51,1,19) for Ba$_{1-x}$K$_x$Fe$_2$As$_2$ ($x$ = 0.00 (a), 0.27 (b)).  Solid lines depict Lorentzian fits.  Spectra at $\bf Q$ = (4.48,0,0) are shifted vertically for ease of viewing 
with the zero level shown by horizontal dashed lines.}
\end{figure}
%=========================================================
$M$Fe$_2$As$_2$ ($M$ = Ba, Ca, or Sr) is a parent compound of Fe-based superconductors.  
A maximum $T_c$ of 38 K is observed for 
Ba$_{1-x}$K$_x$Fe$_2$As$_2$ ($x$ = 0.40) \cite{RotterBaK-2008a}.  
Undoped $M$Fe$_2$As$_2$ shows a tetragonal-to-orthorhombic structural phase transition 
accompanied by the appearance of antiferromagnetic long-range ordering, which can be attributed to the nesting of the Fermi surface along (0.5,0.5,0) in tetragonal notation.  
Although its $T_c$ is lower than the maximum $T_c$ of the $R$FeAsO ($R$ = rare-earth) system, 
$M$Fe$_2$As$_2$ offers the advantages of a simpler crystal structure and 
the availability of large single crystals.  
Both systems share layers formed by FeAs$_4$ tetrahedrons.  
Between these FeAs layers, $R$FeAsO exhibits blocks consisting of two atoms, i.e., 
$R$ and O, while $M$Fe$_2$As$_2$ exhibits blocks consisting of only one atom, i.e., $M$ (Fig. \ref{structure}(a)).  
The simpler crystal structure of $M$Fe$_2$As$_2$ facilitates phonon analysis for 
deriving the nature of superconducting FeAs layers.  
In this study, we therefore analyze the phonon dispersion of Ba$_{1-x}$K$_x$Fe$_2$As$_2$ by inelastic X-ray scattering measurement on single crystals.  

\section{Experimental Procedure}

Single crystals of Ba$_{1-x}$K$_x$Fe$_2$As$_2$ were grown by the self-flux method.  
For the $x$ = 0.00 sample, the FeAs precursor was prepared from Fe and As at 900 $^\circ$C for 10 h in an evacuated atmosphere and then mixed with Ba in the atomic ratio Ba:FeAs = 1:4.  
For the K doped sample, the starting materials of Ba, K, Fe, and As were mixed in the atomic ratio 0.6:0.4:2:2.  
All mixtures were placed in a BN-coated quartz or Al$_2$O$_3$ crucible and sealed in a quartz tube.  
For the $x$ = 0.00 sample, the tube was heated to 1140 $^\circ$C, 
whereas for the K doped sample, the tube was preheated to 600 $^\circ$C, maintained at that temperature for 10 h, and then heated to 1140 $^\circ$C.  
All samples were maintained at the maximum temperature for 10 h and cooled to 1040 $^\circ$C at a rate of 0.4-1 $^\circ$C/h, followed by decanting the flux.  

The compositions of the single crystals were confirmed by lattice-constant measurements 
with a laboratory X-ray diffractometer (RINT-1000, RIGAKU) at room temperature.  
The lattice constant of the K doped sample was $c$ = 13.22 \AA, corresponding to a composition of $x$ = 0.27 \cite{RotterBaK-2008b}.
The $T_c$ of the K doped single crystal was measured
using a SQUID magnetometer at a magnetic field of 10 Oe after zero-field cooling.  
The $T_c$ onset for the $x$ = 0.27 sample is 37 K, close to the maximum $T_c$ 
of the Ba$_{1-x}$K$_x$Fe$_2$As$_2$ system.  

Inelastic X-ray scattering measurements were carried out using 
synchrotron radiation at BL35XU in SPring-8 \cite{Baron2000}.  
The incident X-ray energy was fixed at 21.75 keV using the Si(11,11,11) reflection as 
a monochromator in the backscattering setup.  
The energy resolution was about $\Delta$E = 1.5 meV.  
Single crystals were mounted on a thin carbon fiber attached to a thin glass capillary.  
All measurements were conducted at room temperature.  

Phonon dispersion was analyzed using a Born-von K\'{a}rm\'{a}n force-constant model.  
The longitudinal and transverse force constants of 11 atomic pairs were chosen as fitting parameters (Table \ref{force constant} and Fig. \ref{structure}(a)), and the calculated energies were fitted to the measured data.  
For the $x$ = 0.27 sample, we assumed a virtual atom with an averaged atomic mass 
corresponding to the mixed occupation of this site by Ba and K.  

%=========================================================
\begin{figure*}
\begin{center}
\includegraphics[width=2\columnwidth]{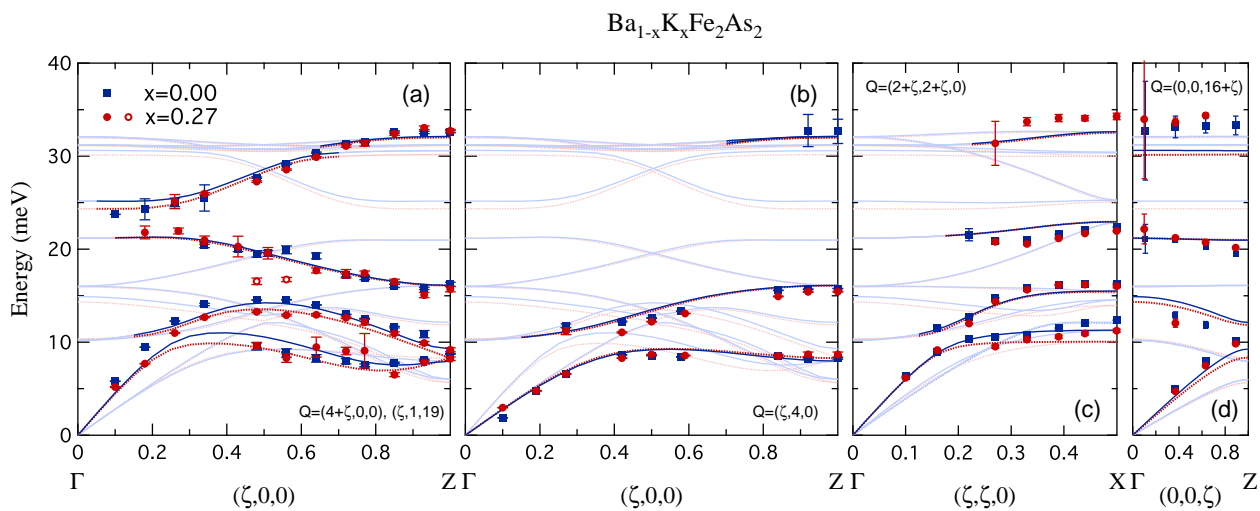}
\caption{\label{dispersion} Phonon dispersion curves of Ba$_{1-x}$K$_x$Fe$_2$As$_2$ ($x$ = 0.00, 0.27).   Solid and dashed lines depict results of calculations based on the Born-von K\'{a}rm\'{a}n model for $x$ = 0.00 and 0.27, respectively.  Light-colored lines depict modes that are unobservable with the chosen scattering vectors.  (a) Phonon modes along ($\zeta$,0,0) observed at $\bf Q$ = (4+$\zeta$,0,0) and ($\zeta$,1,19), (b) at $\bf Q$ = ($\zeta$,4,0), (c) dispersion along ($\zeta$,$\zeta$,0) at $\bf Q$ = (2+$\zeta$,2+$\zeta$,0) and (d) dispersion along (0,0,$\zeta$)  at $\bf Q$ = (0,0,16+$\zeta$).  Squares and circles depict results of measurements for $x$ = 0.00 and 0.27, respectively.}
\end{center}
\end{figure*}
%=========================================================
%=========================================================
\begin{table}
\caption{\label{force constant} Interatomic force constants obtained from analysis based on the Born-von K\'{a}rm\'{a}n model.  Only the force constants of (Ba,K)-Fe and (Ba,K)-(Ba,K) pairs are reduced under K doping.}
%\begin{ruledtabular}
\begin{tabular}{c c c c c c}\hline \hline
      &   Pair &\multicolumn{2}{c}{Longitudinal}&\multicolumn{2}{c}{Transverse}\\
      &   &\multicolumn{2}{c}{force constant}&\multicolumn{2}{c}{force constant}\\
      &          &\multicolumn{2}{c}{(mdyn/\AA)}&\multicolumn{2}{c}{(mdyn/\AA)}\\   \hline
   &     &  $x$=0.00  &  $x$=0.27  &  $x$=0.00  &  $x$=0.27  \\
 $f_1$ & Fe-As &  0.70  &  0.70  &  0.03  &  0.03  \\
 $f_2$ & Fe-Fe &  0.13  &  0.13  &  0.005  &  0.005  \\
 $f_3$ & (Ba,K)-As &  0.08  &  0.08  &  0.002  &  0.002  \\
 $f_4$ & As-As &  0.02 &  0.02  &  0.00  &  0.00  \\
 $f_5$ & (Ba,K)-Fe &  0.10  &  0.07  &  0.003  &  0.002  \\
 $f_6$ & As-As &  0.03  &  0.03  &  0.001  &  0.001  \\
 $f_7$ & As-As &  0.04  &  0.04  &  0.002  &  0.002  \\
 $f_8$ & (Ba,K)-(Ba,K) &  0.08  &  0.03  &  0.003  &  0.002  \\
 $f_9$ & Fe-Fe &  0.03  &  0.03  &  0.002  &  0.002  \\
 $f_{10}$ & Fe-As &  0.03  &  0.03  &  0.003  &  0.003  \\
 $f_{11}$ & (Ba,K)-(Ba,K) &  0.02  &  0.01  &  0.002  &  0.001  \\
 \hline \hline
\end{tabular}
%\end{ruledtabular}
\end{table}
%=========================================================
\section{Results and Discussion}

Figure \ref{spectrum-nest} shows the energy spectra of the $x$ = 0.00 and 0.27 samples 
at the scattering vector $\bf Q$ = (2.5,2.5,0), where the wave vector $\bf q$ = (0.5,0.5,0) 
is the same as the nesting vector of the Fermi surface.  
The solid lines depict Lorentzian fits.  
Comparing the $x$ = 0.00 and 0.27 samples, 
the well-defined peak around $E$ = 12 meV softens about 1 meV and the linewidth broadens 
upon K doping.  
For the other peaks, the doping-induced energy shifts amount to less than 0.5 meV.  
These observed phonon spectra are almost consistent with previous data reported in ref. 13, 
except for the softening and broadening.  
Note that the phonon mode at $E$ = 22 meV, which has been predicted to couple strongly with 
the static magnetic moment \cite{Mittal2009, Reznik2009}, 
exhibits only a weak K doping dependence, 
although magnetic long-range ordering of $x$ = 0.00 disappears with K doping.  

To clarify whether the softening and broadening are associated 
with nesting or a magnetic interaction, 
we measured phonon spectra at a different $\bf Q$ position of approximately (4.5,0,0), where $\bf q$ is 
tilted at 45$^\circ$ to the nesting vector (Fig. \ref{spectrum-long}).  
The solid lines depict Lorentzian fits.  
As shown, the well-defined peak around $E$ = 14 meV softens about 1 meV upon K doping 
accompanied by a broadening similar to the observation at $\bf Q$ = (2.5,2.5,0).  
Since similar softening and broadening are observed along different $\bf q$ directions, 
it appears unlikely that these effects are caused by Fermi surface nesting 
or a magnetic interaction.  
A displacement pattern of the well-defined peaks around $E$ = 14 meV is shown in Fig. \ref{structure}(b).  
These modes are characterized by large (Ba,K) amplitudes, 
rendering them very sensitive to K doping.  
The broadening may be due to distortion induced by K doping.  

Phonon modes with energies around $E$ = 16-20 meV at $\bf Q$ = (4.48,0,0) 
also show softening from $E$ = 19.5 to 16.5 meV with K doping (Fig. \ref{spectrum-long}).  
Since the corresponding phonon intensity is weak owing to the small dynamical structure factor 
and disturbing peaks present around $E$ = 14 meV, 
we have confirmed the spectra around $\bf Q$ = (0.5,1,19) where $\bf q$ is identical and 
the lattice-dynamics model predicts a higher intensity for the mode 
around $E$ = 16-20 meV (Fig. \ref{spectrum-trans}).  
The displacement pattern of the corresponding phonon mode is shown in Fig. \ref{structure}(c).  
For the $x$ = 0.00 sample, a well-defined peak is observed at $E$ = 19.5 meV around 
$\bf Q$ = (0.5,1,19), which agrees well with the energy observed around $\bf Q$ = (4.5,0,0) (Fig. \ref{spectrum-trans}(a)).  
On the other hand, for the $x$ = 0.27 sample, the peak observed at $E$ = 16.5 meV 
around $\bf Q$ = (4.5,0,0) is not observable around $\bf Q$ = (0.5,1,19) (Fig. \ref{spectrum-trans}(b)).  
However, a well-defined peak is observed at $E$ = 19.5 meV, which is the same energy 
as that observed for the $x$ = 0.00 sample.  
Therefore, the peak observed at $E$ = 16.5 meV around $\bf Q$ = (4.5,0,0) 
in the doped sample is unlikely to be associated with the $E$ = 19.5 meV signals; 
it appears to be associated with the additional mode at the lower energy appearing in only the doped sample.  

Figure \ref{dispersion} shows the observed and calculated dispersion curves for the $x$ = 0.00 and 0.27 samples.  
The phonon modes shown in Figs. \ref{spectrum-long} and \ref{spectrum-trans} 
are indicated in Fig. \ref{dispersion}(a), and those shown in Fig. \ref{spectrum-nest} are indicated 
in Fig. \ref{dispersion}(c).  
The extra phonon modes appearing around $\bf Q$ = (4.5,0,0) in the $x$ = 0.27 sample are depicted by open squares in Fig. \ref{dispersion}(a).  
Doping-induced phonon softening is observed in the energy range $E$ = 10-15 meV at $\bf Q$ = (4+$\zeta$,0,0) and (2+$\zeta$,2+$\zeta$,0).  
The energies of all the other phonon modes studied are almost identical for the $x$ = 0.00 and 0.27 samples.  
All data, except the extra modes, are reasonably well described by the calculation.  
The obtained interatomic force constants are shown in Table \ref{force constant}.  

The observed phonon softening is reproduced by reducing the 
(Ba,K)-(Ba,K) and (Ba,K)-Fe force constants.  
As shown in Fig. \ref{dispersion}, the calculated energies of the corresponding $x$ = 0.00 phonon modes 
(solid lines) that soften upon doping decrease toward the dashed lines of the $x$ = 0.27 phonon modes.  
The energies of these phonon modes are sensitive to the force constants involving the (Ba,K) atoms, 
since these atoms contribute largely to the corresponding polarization pattern (see Fig. \ref{structure}(b)).  
The smaller (Ba,K)-(Ba,K) and (Ba,K)-Fe force constants of the K doped sample appear to be reasonable
owing to the fact that monovalent K ions exhibit a smaller valency than divalent Ba ions.  

The extra modes observed around $\bf Q$ = (4.5,0,0) and $E$ = 16.5 meV, on the other hand, 
cannot be explained by the Born-von K\'{a}rm\'{a}n harmonic model.  
A plausible explanation for the extra modes is the splitting of lower-
energy modes near $E$ = 13 meV, where (Ba,K) atoms dominate the polarization pattern 
(Fig. \ref{structure}(b)).
The substitution of Ba by K induces significant local disorder due to the difference in ionic radius 
thereby allowing for mode splitting as well as for broadening.  
In addition, the difference in atomic mass by nearly a factor of three may enhance the splitting 
of phonon frequency.  
In fact, in Ba$_{1-x}$K$_x$BiO$_3$ superconducting materials, K doping leads to 
disorder that has been characterized in ref. 24.  
However, we may not rule out the possibility that extra modes in Ba$_{0.73}$K$_{0.27}$Fe$_2$As$_2$ 
arise from electron-phonon coupling.  
Recently, $^{57}$Fe nuclear resonant inelastic scattering measurement has revealed that 
the Fe partial density of states of Ba$_{1-x}$K$_x$Fe$_2$As$_2$ shows softening 
around $E$ = 20 meV with K doping \cite{Tsutsui2009}.  
It could be that the extra modes split rather from the modes located at a higher energy, i.e., 
$E$ = 19.5 meV, owing to electron-phonon coupling.  
NdFeAsO$_{1-y}$F$_y$ powder samples analyzed by inelastic X-ray scattering measurement also show 
phonon softening at a similar energy of $E$ = 20 meV with carrier doping \cite{Tacon2008}.  
Because the present modes around $\bf Q$ = (4.5,0,0) and $E$ = 16.5 meV 
consist mainly of Fe and As atom vibrations (Fig. \ref{structure}(c)), 
it is reasonable for the soft phonon modes in NdFeAsO$_{1-y}$F$_y$ 
to be of comparable energy.  
It is worth noting that As-Fe-As bond angles that can be associated with 
$T_c$ \cite{Lee2008} vibrate in the corresponding phonon modes.  

\section{Conclusions}

We have studied the phonon dynamics of 
Ba$_{1-x}$K$_x$Fe$_2$As$_2$ ($x$ = 0.00, 0.27) by inelastic X-ray scattering measurement.  
K doping induces phonon softening and broadening for
modes in the energy range $E$ = 10-15 meV. These modes are
characterized by large (Ba,K) vibrations.  Born-von
K\'{a}rm\'{a}n analysis indicates that the softening results from
a reduction in interatomic force constants. The broadening can be
a consequence of the local disorder induced by K substitution.
Extra modes observed around $\bf q$ = (0.5,0,0) and $E$ = 16.5 meV
for the $x$ = 0.27 sample appear to arise from the phonon splitting due to
either the local disorder or the electron-phonon interaction.

\begin{acknowledgments}
We thank I. Hase, N. Takeshita, K. Miyazawa, H. Nakamura, M. Machida, A. Komarek, 
T. Hasegawa, and M. Udagawa for valuable discussions.  
The present experiments were performed under the approval of JASRI (Proposal No. 2008B1381 and 2009A1146).
This work was supported by a Grant-in-Aid for Scientific Research on Innovative 
Areas ``Heavy Electrons" (No. 20102005) from MEXT, 
Japan, and performed under the interuniversity cooperative
research program of the Institute for Materials Research, Tohoku University.  
Work at Cologne University was supported by the DFG through SFB 608.
\end{acknowledgments}


\begin{thebibliography}{9}
\bibitem{Kamihara2008} Y. Kamihara, T. Watanabe, M. Hirano, and H. Hosono: J. Am. Chem. Soc. {\bf130} (2008) 3296.
\bibitem{Ren-F-2008} Z. A. Ren, W. Lu, J. Yang, W. Yi, X. L. Shen, Z. C. Li, G. C. Che, X. L. Dong, L. L. Sun, F. Zhou, and Z. X. Zhao: Chin. Phys. Lett. {\bf25} (2008) 2215.
\bibitem{Kito2008} H. Kito, H. Eisaki, and A. Iyo: J. Phys. Soc. Jpn. {\bf77} (2008) 063707.
\bibitem{Ren-O-2008} Z. A. Ren, G. C. Che, X. L. Dong, J. Yang, W. Lu, W. Yi, X. L. Shen, Z. C. Li, L. L. Sun, F. Zhou, and Z. X. Zhao: Europhys. Lett. {\bf83} (2008) 17002.
\bibitem{Miyazawa2009} K. Miyazawa, K. Kihou, P. M. Shirage, C. H. Lee, H. Kito, H. Eisaki, and A. Iyo: J. Phys. Soc. Jpn. {\bf78} (2009) 034712.
\bibitem{Litvinchuk2008} A. P. Litvinchuk, V. G. Hadjiev, M. N. Iliev, Bing Lv, A. M. Guloy, and C. W. Chu: Phys. Rev. B {\bf78} (2008) 060503(R).
\bibitem{Fukuda2008} T. Fukuda, A. Q. R. Baron, S. Shamoto, M. Ishikado, H. Nakamura, M. Machida, H. Uchiyama, S. Tsutsui, A. Iyo, H. Kito, J. Mizuki, M. Arai, H. Eisaki, and H. Hosono: J. Phys. Soc. Jpn. {\bf77} (2008) 103715.
\bibitem{Christianson2008} A. D. Christianson, M. D. Lumsden, O. Delaire, M. B. Stone, D. L. Abernathy, M. A. McGuire, A. S. Sefat, R. Jin, B. C. Sales, D. Mandrus, E. D. Mun, P. C. Canfield, J. Y. Y. Lin, M. Lucas, M. Kresch, J. B. Keith, B. Fultz, E. A. Goremychkin, and R. J. McQueeney: Phys. Rev. Lett. {\bf101} (2008) 157004.
\bibitem{Mittal2008} R. Mittal, Y. Su, S. Rols, T. Chatterji, S. L. Chaplot, H. Schober, M. Rotter, D. Johrendt, and Th. Brueckel: Phys. Rev. B {\bf78} (2008) 104514.
\bibitem{Zbiri2009} M. Zbiri, H. Schober, M. R. Johnson, S. Rols, R. Mittal, Y. Su, M. Rotter, and D. Johrendt: Phys. Rev. B {\bf79} (2009) 064511.
\bibitem{Mittal2009} R. Mittal, L. Pintschovius, D. Lamago, R. Heid, K-P. Bohnen, D. Reznik, S. L. Chaplot, Y. Su, N. Kumar, S. K. Dhar, A. Thamizhavel, and Th. Brueckel: Phys. Rev. Lett. {\bf102} (2009) 217001.
\bibitem{Tacon2008} M. Le Tacon, M. Krisch, A. Bosak, J.-W. G. Bos, and  S. Margadonna: Phys. Rev. B {\bf78} (2008) 140505(R).
\bibitem{Reznik2009} D. Reznik, K. Lokshin, D. C. Mitchell, D. Parshall, W. Dmowski, D. Lamago, R. Heid, K.-P. Bohnen, A. S. Sefat, M. A. McGuire, B. C. Sales, D. G. Mandrus, A. Subedi, D. J. Singh, A. Alatas, M. H. Upton, A. H. Said, A. Cunsolo, Yu. Shvyd'ko, and T. Egami: arXiv:0908.4359.
\bibitem{Boeri2008} L. Boeri, O. V. Dolgov, and A. A. Golubov: Phys. Rev. Lett. {\bf101} (2008) 026403.
\bibitem{Noffsinger2009} J. Noffsinger, S. G. Louie, M. L. Cohen, and F. Giustino: Phys. Rev. Lett. {\bf102} (2009) 147003.
\bibitem{Lee2008} C. H. Lee, A. Iyo, H. Eisaki, H. Kito, M. T. Fernandez-Diaz, T. Ito, K. Kihou, H. Matsuhata, M. Braden, and K. Yamada: J. Phys. Soc. Jpn. {\bf77} (2008) 083704.
\bibitem{Singh2008} D. J. Singh and M.-H. Du: Phys. Rev. Lett. {\bf100} (2008) 237003.
\bibitem{Kuroki2009} K. Kuroki, H. Usui, S. Onari, R. Arita, and H. Aoki: Phys. Rev. B {\bf79} (2009) 224511.
\bibitem{Yildirim2009} T. Yildirim: Phys. Rev. Lett. {\bf102} (2009) 037003.
\bibitem{Eschrig2009} H. Eschrig: arXiv:0804.0186.
\bibitem{RotterBaK-2008a} M. Rotter, M. Tegel, and D. Johrendt: Phys. Rev. Lett. {\bf101} (2008) 107006.
\bibitem{RotterBaK-2008b} M. Rotter, M. Pangerl, M. Tegel, and D. Johrendt: Angew. Chem. Int. Ed. {\bf 47} (2008) 7949.
\bibitem{Baron2000} A. Q. R. Baron, Y. Tanaka, S. Goto, K. Takeshita, T. Matsushita, and T. Ishikawa: J. Phys. Chem. Solids {\bf61} (2000) 461.
\bibitem{braden-BaKBiO} M. Braden, W. Reichardt, E. Elkaim, J. P. Lauriat, S. Shiryaev, and S. N. Barilo: Phys. Rev. B {\bf 62} (2000) 6708.
\bibitem{Tsutsui2009} S. Tsutsui, C. H. Lee, C. Tassel, Y. Yoshida, Y. Yoda, K Kihou, A. Iyo, and H. Eisaki: submitted to J. Phys. Soc. Jpn.
\end{thebibliography}
\end{document}